\documentclass[%
preprint,
amsmath,amssymb,
prl,
longbibliography,
superscriptaddress,
]{revtex4-1}
\usepackage{amssymb}
\usepackage{amsmath}
\usepackage{epsfig}
\usepackage{epstopdf}
\usepackage{bm}
\usepackage{graphicx,epsfig}
\usepackage{mathrsfs}
\usepackage{dcolumn}
\usepackage{color}
\usepackage{natbib}
\usepackage{CJK}
\usepackage{color}
\usepackage{graphicx}
\usepackage{dcolumn}
\usepackage{bm}

\usepackage{lineno}

\begin{document}
\title{Simulating graphene dynamics in one-dimensional modulated ring array with synthetic dimension}

\author{Danying Yu}\thanks{These authors contribute equally to this work.}
\affiliation{State Key Laboratory of Advanced Optical Communication Systems and Networks, School of Physics and Astronomy, Shanghai Jiao Tong University, Shanghai 200240, China}

\author{Guangzhen Li}\thanks{These authors contribute equally to this work.}
\affiliation{State Key Laboratory of Advanced Optical Communication Systems and Networks, School of Physics and Astronomy, Shanghai Jiao Tong University, Shanghai 200240, China}

\author{Meng Xiao}
\affiliation{Key Laboratory of Artificial Micro- and Nano-Structures of Ministry of Education and School of Physics and Technology, Wuhan University, Wuhan 430072, China}

\author{Da-Wei Wang}
\affiliation{Interdisciplinary Center for Quantum Information, State Key Laboratory of Modern Optical Instrumentation, and Zhejiang Province Key Laboratory of Quantum Technology and Device, Department of Physics, Zhejiang University, Hangzhou 310027, China}

\author{Yong Wan}
\affiliation{National Institute of Metrology, Beijing 102200, China}

\author{Luqi Yuan}
\email{yuanluqi@sjtu.edu.cn}
\affiliation{State Key Laboratory of Advanced Optical Communication Systems and Networks, School of Physics and Astronomy, Shanghai Jiao Tong University, Shanghai 200240, China}

\author{Xianfeng Chen}
\affiliation{State Key Laboratory of Advanced Optical Communication Systems and Networks, School of Physics and Astronomy, Shanghai Jiao Tong University, Shanghai 200240, China}
\affiliation{Shanghai Research Center for Quantum Sciences, Shanghai 201315, China}
\affiliation{Jinan Institute of Quantum Technology, Jinan 250101, China}
\affiliation{Collaborative Innovation Center of Light Manipulations and Applications, Shandong Normal University, Jinan 250358, China}


\begin{abstract}
A dynamically-modulated ring system with frequency as a synthetic dimension has been shown to be a powerful platform to do quantum simulation and explore novel optical phenomena. Here we propose synthetic honeycomb lattice in a one-dimensional ring array under dynamic modulations, with the extra dimension being the frequency of light. Such system is highly re-configurable with modulation. Various physical phenomena associated with graphene including Klein tunneling, valley-dependent edge states, effective magnetic field, as well as valley-dependent Lorentz force can be simulated in this lattice, which exhibits important potentials for manipulating photons in different ways. Our work unveils a new platform for constructing the honeycomb lattice in a synthetic space, which holds complex functionalities and could be important for optical signal processing as well as quantum computing.
\end{abstract}

\maketitle

\section*{Introduction}
The honeycomb lattice, with the same geometry as the graphene \cite{32,33}, attracts great interest in condensed matter physics \cite{c1,c4} and photonics \cite{b26,b15}. Rich physical phenomena have been reported in photonic honeycomb lattices, taking advantages of properties of Dirac points and the valley degree of freedom \cite{Ezawa2013,a20,c5,21,c6,c7,41,42}  and showing excellent platform for studying topological photonics \cite{a33}, which also have potential applications in the interface of nonlinear optics \cite{a49} and quantum optics \cite{a50}, pointing towards topological fiber \cite{a14}, topological laser \cite{a47}, as well as edge and gap solitons \cite{a23}. Different platforms have been achieved to construct photonic honeycomb lattices, such as waveguide arrays \cite{b24}, on-chip silicon photonics \cite{b21}, semi-conductor microcavities \cite{Segev2018}, and metamaterials \cite{a21}. However, one notes that the reconfigurability and feasibility of a system are attractive for satisfying various experimental requirements and different applications, as well as relaxing the fabrication constrains, which are naturally limited in the current honeycomb systems due to their fixed configurations or structures after fabrication. Therefore, it is of significant importance to find an alternative platform, which is experimentally feasible and holds reconfigurability.

Here, we propose the construction of a highly re-configurable honeycomb lattice in a synthetic space in a modulated ring resonator system. A ring resonator under dynamic modulation has been found to be capable for creating a synthetic dimension along the frequency axis of light \cite{54}, which together with spatial dimensions, a variety of physical implementations have been suggested with two or more dimensions \cite{54,53,57,b3,b4,b5,b6}. We show that a one-dimensional ring resonator array composed by two types of resonators under proper dynamic modulations supports a two-dimensional honeycomb lattice in a synthetic space including both spatial and frequency dimensions. Different physics associated with the photonic graphene can be simulated in this unique platform, such as Klein tunneling \cite{a20}, valley-dependent edge states \cite{21}, topological edge states with the effective magnetic field \cite{Yin2011}, and valley-dependent Lorentz force \cite{42}. The modulated ring systems can be achieved in either fiber-based system \cite{55,51,b7,88} or on-chip lithium niobate resonator \cite{b8}, which brings our proposal to a flexible experimental setup with state-of-art technologies in bulk optics or integrated photonics. Our work not only broadens the current research on synthetic dimensions in photonics \cite{46,b9}, but also enriches quantum simulations with topological photonics \cite{47,b10}, which shows potential applications in optical signal processing \cite{c13,c10} and quantum computations \cite{c15,c14}.

\section*{Results}
\textbf{Model.}
We begin with considering a pair of ring resonators (labeled as A and B) with the same circumference $L$ undergoing dynamic modulation [see Fig.~1(a)]. The central resonant frequencies of ring A and ring B are set at $\omega_0$ and $\omega_0-\Omega/2$, respectively. In the absence of group velocity dispersion, the frequency of the $m^\mathrm{th}$ resonant mode in ring A (ring B) is $\omega_{m,\mathrm{A}}=\omega_0+m\Omega$ ($\omega_{m,\mathrm{B}}=\omega_0-\Omega/2+m\Omega$), where $\Omega=2\pi v_g/L$ is the free spectral ranges (FSR) with $v_g$ being the group velocity inside both rings. We place electro-optic modulators (EOM) inside two rings, with modulation frequency $\Omega/2$ and modulation phase $\phi$. A synthetic frequency dimension with the effective hopping amplitude $g$ can be constructed with spaced frequency $\Omega/2$ in the frequency axis of light, where modes supported by rings A and B are labeled by $a$ and $b$, respectively. With the building block for constructing the one-dimensional synthetic frequency dimension in a pair of rings, we can then use it to construct a synthetic honeycomb lattice in a one-dimensional array of pairs of rings shown in Fig.~1(b) [see Methods]. The ring array consists groups of rings ($n=1,2,...,N$), each of which contains two pairs of rings with different combinations, i.e., AB (labeled as $\alpha=1$) and BA (labeled as $\alpha=2$), respectively. We write the Hamiltonian of the system under the first-order approximation:
\begin{eqnarray}
H_r&=&\sum_{m,n}[\kappa(a_{n,m,1}^\dagger a_{n-1,m,2}+ b_{n,m,2}^\dagger b_{n,m,1})\nonumber\\
&&+g_{2n-1} (a_{n,m,1}^\dagger b_{n,m,1}e^{i\phi_{2n-1}}+a_{n,m,1}^\dagger b_{n,m+1,1}e^{-i\phi_{2n-1}})\nonumber\\
&&+g_{2n}(b_{n,m,2}^\dagger a_{n,m-1,2}e^{i\phi_{2n}}+b_{n,m,2}^\dagger a_{n,m,2}e^{-i\phi_{2n}})]+h.c.,
\end{eqnarray}
where $a_{n,m,\alpha}^\dagger$ ($a_{n,m,\alpha}$) and $b_{n,m,\alpha}^\dagger$ ($b_{n,m,\alpha}$) are corresponding creation (annihilation) operators, and $\kappa$ is the evanescent-wave coupling strength between two rings at the same type. The Hamiltonian in Eq.~(1) therefore supports a two-dimensional synthetic honeycomb lattice [see Fig.~1(c)].

\begin{figure}[htbp]
\centering
\includegraphics[width=0.85\textwidth ]{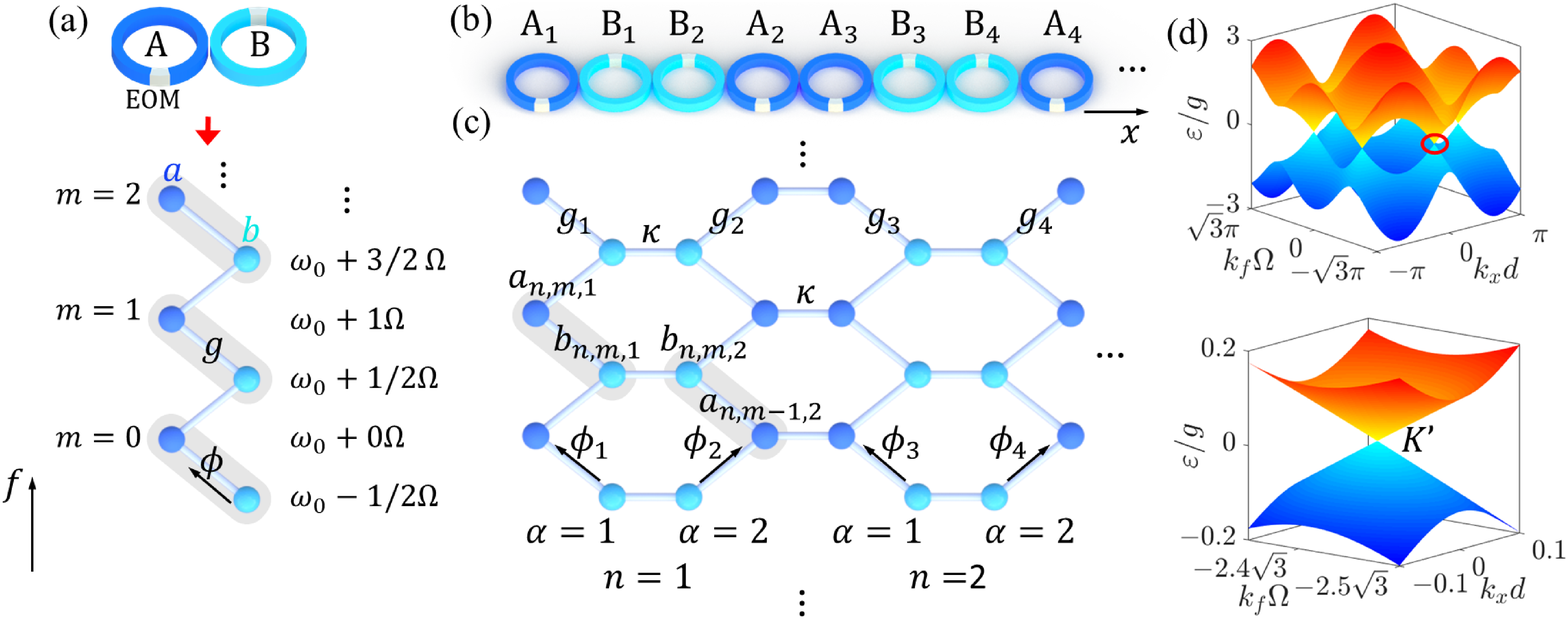}
\caption{\textbf{Construction of synthetic honeycomb lattice.} (a) Rings A and B are modulated by EOM, which supports a set of resonant modes $a$ (blue circle) and $b$ (cyan circle). (b) A one-dimensional array of ring resonators composed by groups of pairs of rings in (a). (c) The system in (b) can be mapped into a honeycomb lattice in a synthetic space including one spatial dimension and one frequency dimension. (d) The band structure (upper panel) and the zoomed-in Dirac cone at $K'$ point (lower panel) of the honeycomb lattice in (c).}
\end{figure}

The honeycomb lattice is constructed in a synthetic space including the spatial ($x$) and frequency ($f$) dimensions. Different from conventional photonic honeycomb lattice in real space \cite{a20,21} that depends greatly on apparent geometry, the synthetic honeycomb lattice in Eq. (1) is dependent on both couplings between rings ($\kappa$) and modulations ($g_i$). Therefore, without loss of generality, we label the distance between two sites in the synthetic lattice with same types as $d$ and the distance between two sites with different types as $d/2$ along the $x$-axis in the later plots of field patterns. The flexible choice of hopping amplitude $g_i$ and phase $\phi_i$ provides the powerful reconfigurability towards different physical phenomena in quantum simulations. We first consider $\kappa = g_i=g$ and $\phi_i=0$, and the synthetic lattice holds the unit cell with the translation symmetry including two frequency modes $a$ and $b$ in two rings. The band structure of this honeycomb lattice in the first Brillouin zone can be plotted in the $k_x$-$k_f$ space, where $k_x$ and $k_f$ are wave vectors reciprocal to the spatial ($x$) and frequency ($f$) axes, respectively. Dirac point $K$ and $K'$ at $(k_x,k_f)=(0,\pm 4\pi/3\Omega)$ can be seen in the band structure in Fig.~1(d), where the zoomed-in Dirac cone shows linear dispersion near $K'$ point. In the following, we will show the capability for achieving different phenomena associated to the honeycomb lattice with the current platform.

\textbf{Klein tunneling.}
Klein tunneling, as an intriguing phenomenon in physics exploring a particle passing through a barrier higher than its energy, has experienced great interest in different platforms including graphene and photonic/phononic crystals \cite{a20,c5}. To demonstrate such physics in the synthetic honeycomb lattice in Fig.~1(c), we couple the first (A$_1$) and the last (A$_{2N}$) rings with an external waveguide such that a periodic condition along the $x$-axis can be naturally created. Such a design forms a carbon-nanotube-like shape [see Fig.~2(a)]. In simulation, we consider $N=12$, which corresponds to $48$ rings ranging inside $x$$\in$$[0, 35d]$, and modes in the region $f$$\in$$[-8\Omega, 47.5\Omega]$. We first excite the honeycomb lattice by injecting an input plane wave with distribution of $s_1=e^{-(f-f_0)^2/\Delta^2}e^{ik_x(x-x_0)+ik_f(f-f_0)}$ to excite the initial wave packet of the field shown in Fig.~2(b), where $f_0=39\Omega$, $x_0=17.5d$, and $\Delta=8.95\Omega$, and collect signals through external waveguides to readout the evolution of the field in the synthetic honeycomb lattice [see Methods]. We set the $k$-vector in $s_1$ in the vicinity of the Dirac point $K$ by using $(k_x,k_f)=(0,4.04\pi/3\Omega)$. The corresponding frequency detuning of the source is $\varepsilon=0.036g$, which falls in the linear dispersion region in the Dirac cone as indicated in Fig.~1(d) and gives the initial velocity of the wave packet along the negative $k_f$ direction. The simulation verifies this feature, that the field propagates to the bottom of the synthetic honeycomb lattice at $t=42g^{-1}$ as shown in Fig.~2(c) without distortion in shape, since there is no barrier in this case.

Based on the calculation in Fig.~2(c), we set an artificial square-shape barrier in the middle range of the frequency dimension in the synthetic honeycomb lattice as shown in Fig.~2(d), which in principal can be achieved by adding on-site potential terms $Va_{n,m,\alpha}^\dagger a_{n,m,\alpha} $ and $Vb_{n,m,\alpha}^\dagger b_{n,m,\alpha}$ in the range $f$$\in$$[12,35.5]\Omega$ with $V=0.05g$. Although $\varepsilon<V$, the field experiences an effective lift in its energy in the barrier region, but still sticks to the linear region of the Dirac cone, so the initial negative group velocity does not change during this process [see the upper panel in Fig.~2(d)]. This is indeed verified in the simulation in Fig.~2(d), which shows that the field propagates to the bottom of the synthetic honeycomb at $t=42g^{-1}$ without significant change in the field distribution.

\begin{figure}[htbp]
\centering
\includegraphics[width=0.65\textwidth ]{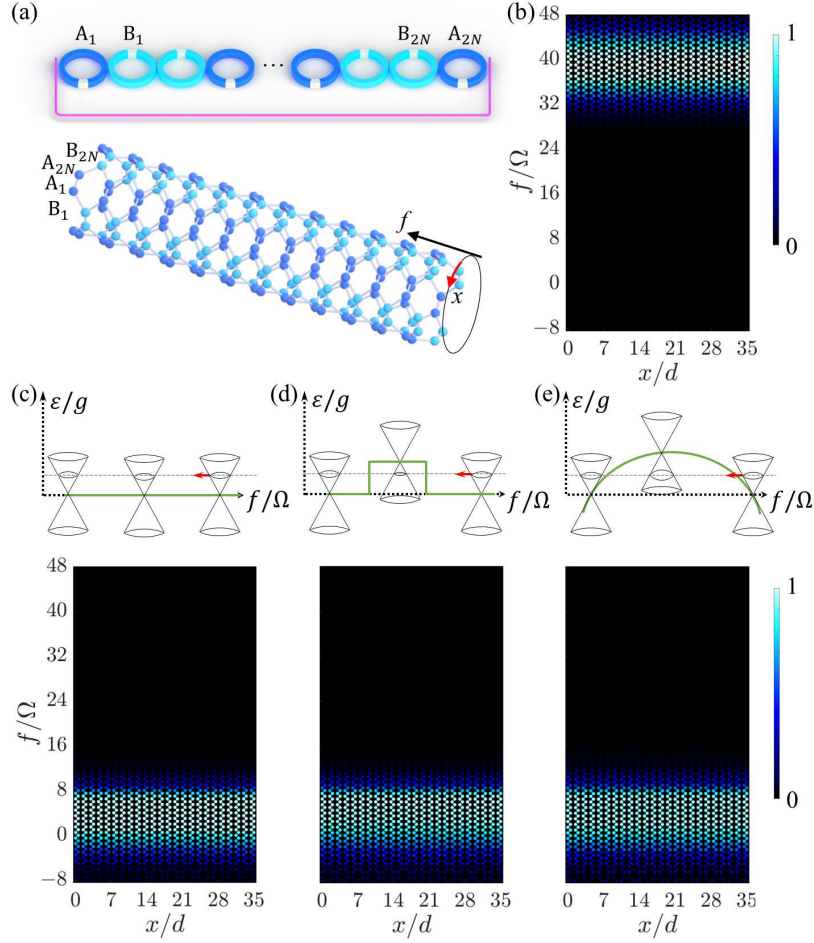}
\caption{\textbf{Klein tunneling.} (a) An external waveguide connects the first (A$_1$) and last (A$_{2N}$) ring resonator together, which forms an effective carbon-nanotube-like honeycomb lattice in the synthetic space with $x$ direction being periodic. (b) The intensity distribution in the synthetic lattice of the initial plane-wave excitation at $t=0$.  Intensity distributions in the synthetic lattice in simulations at $t_f=42g^{-1}$ with (c) no barrier, (d) square-shape potential barrier, and (e) parabola-shape potential barrier (green lines). The upper panels are cartoons showing the excited wave packet in the Dirac cone moving towards the negative direction along the frequency dimension with unchanged $\varepsilon$ (dashed gray line). The red arrows present the direction of velocity.}
\end{figure}

The artificial square-shape barrier along the frequency axis of light is not easy to be constructed in the frequency dimension. In waveguide that composes the ring, there exists the group velocity dispersion that can introduce on-site potentials at modes with different resonant frequencies \cite{57,g1}. We now take the dispersion back into the consideration only in this part, with the zero dispersion point at $f_1=23.75\Omega$. The modulation frequency is chosen to be resonant with the frequency spacing between modes near $f=8.5\Omega$ (and $39\Omega$), which results in an effective parabola-shape potential barrier $V(f)=0.05g+2D(f/\Omega-f_1/\Omega)^2$ with $D=-1.075\times 10^{-4}g$ [see the upper panel in Fig.~2(e)]. Such the group velocity dispersion can be designed by the waveguide-structure engineering \cite{90}. Note here the maximum value of the barrier is $0.05g>\varepsilon$. Different from the previous case in Fig.~2(d) that the field experiences a sudden change in the synthetic $k$-space, when the wave packet of the field propagates towards bottom of the synthetic lattice, it experiences a gradual change inside the Dirac cone. Yet, the distribution of the wave packet still remain largely unchanged at $t=42g^{-1}$, as one sees in the bottom panel of Fig.~2(e).  The propagation of the field is distorted if one uses a larger barrier, i.e., beyond the limit of the Dirac cone. As long as the potential change is limited inside the linear region near the Dirac point, the constructed synthetic honeycomb lattice supports the Klein tunneling along the frequency axis of light.

\textbf{Valley-dependent edge state.}
Valley-dependent photonic phenomena recently attract a broad interest in photonics for providing valley degree of freedom, which offers a new possibility to manipulate light and finds important applications in optical encoding and enlarging the optical information capacity \cite{21}. Here we show the existence of valley-dependent edge states in the synthetic space. Lengths of rings are carefully adjusted to introduce the required effective on-site potential in each ring, which breaks the inversion symmetry of the synthetic lattice and lifts the degeneracy at Dirac points $K$ and $K'$.

We consider 12 pairs of rings (24 rings), each of which has a different circumference $L'$ close to the reference length $L$. The offset in length leads to a slightly shifted resonant frequencies in each ring, which leads to the effective on-site potential on each column site in the synthetic lattice as shown in Fig.~3(a) [see Methods]. As one can see, we consider that there are on-site potentials $\pm U_0$ alternatively in each ring except for the middle two rings having $-U_0$ forming the artificial domain wall. The band structures with an infinite frequency dimension and finite rings can be calculated. Figs.~3(b) and 3(c) plot the projected band structures with potentials $U_0=0$ and $U_0=0.5g$, respectively. One sees that the Dirac point at $K$ ($K'$) with $k_f=4\pi/3\Omega$ ($k_f=2\pi/3\Omega$) is open and valley-dependent edge states are shown when effective on-site potentials are added.

\begin{figure}[htbp]
\centering
\includegraphics[width=0.65\textwidth ]{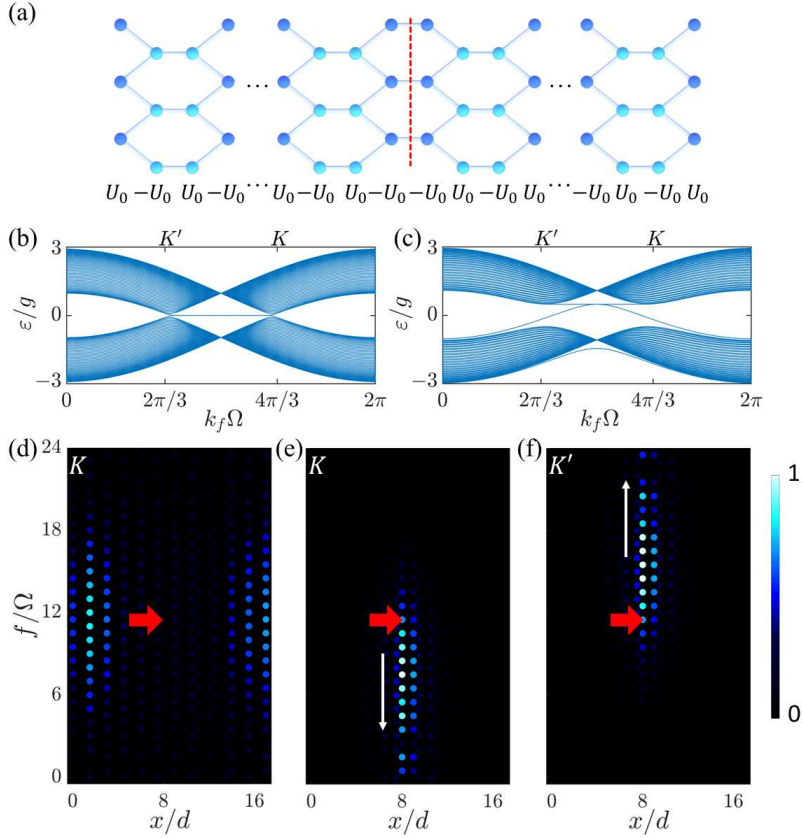}
\caption{\textbf{Valley-dependent edge state.} (a) Effective on-site potentials are alternatively applied in the synthetic lattice, with an artificial domain wall in the middle two columns of synthetic lattices (labelled in red dashed line). The projected band structure in finite rings with (b) $U_0=0$, and (c) $U_0=0.5g$. The simulated intensity distributions at $t=10g^{-1}$ under conditions: (d) $U_0=0$, and $k_f=4\pi/3\Omega$; (e) $U_0=0.5g$, and $k_f=4\pi/3\Omega$; (f) $U_0=0.5g$, and $k_f=2\pi/3\Omega$. Red arrows indicate the excitation at the $12^\mathrm{th}$ ring with the central frequency mode at $f_2=11.5\Omega$.}
\end{figure}

To verify edge states in two valleys in the synthetic honeycomb lattice, we further assume that there are boundaries at the frequency dimension, which can be achieved either by adding auxiliary rings to knock out certain modes at particular frequency \cite{h1} or by designing a sharp change in the group velocity dispersion of the waveguide that composes the ring \cite{54}. Therefore, a lattice with the range of $x$$\in$$[0,17d]$ and $f$$\in$$[0,23.5\Omega]$ is considered in simulations. We excite the $12^\mathrm{th}$ ring on the artificial domain wall by a source field which has the Gaussian spectrum: $s_2=e^{ik_f\cdot (f-f_2)}\cdot e^{-(f-f_2)^2/\Delta^2}$, with $f_2=11.5\Omega$ and $\Delta=8.49\Omega$. $k_f$ here indicates the relative phase information for different frequency components in the source. We first choose $k_f=4\pi/3\Omega$, which excites states near $K$ point, and the simulation results at $t=10g^{-1}$ with $U_0=0$ and $U_0=0.5g$ are plotted in Figs.~3(d) and 3(e), respectively. One can see that when effective potentials are zero and the $K$ point is degenerate, the field leaves the domain wall and spreads into left and right sides of the synthetic lattice. On the other hand, when there are non-zero potentials shown in Fig.~3(a), the one-way edge state at the $K$ valley is excited and propagates towards lower frequency components with most of energy concentrated in the middle two rings on the artificial domain wall. Moreover, if we choose $k_f=2\pi/3\Omega$ in the source to excite the edge state near the $K'$ point, the field experience unidirectionally up-conversion in the middle two columns of the synthetic lattice [see Fig.~4(f)], which shows the possibility of achieving valley-dependent edge states in the synthetic lattice.

\begin{figure}[htbp]
\centering
\includegraphics[width=0.75\textwidth ]{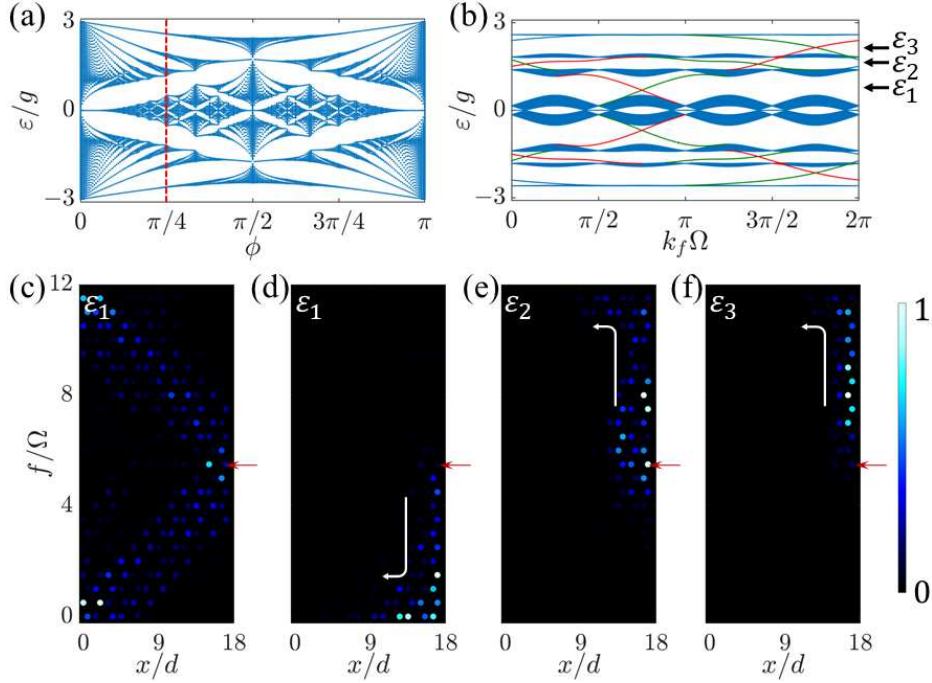}
\caption{\textbf{Topological edge states with effective magnetic field.} (a) The projected band structure of the honeycomb lattice in Fig.~1(c) along $\phi$. (b) The projected band structure in finite rings along $k_f$ with $\phi=\pi/4$. Bands labelled in red (green) represents edge states on the right (left) boundary of the synthetic lattice. (c) The simulated intensity distribution at $t=20g^{-1}$ with the input field having frequency shift $\omega_s=\varepsilon_1$ for the case $\phi=0$. Simulated  intensity distributions at $t=20g^{-1}$  for the case $\phi=\pi/4$ with the input field having frequency shift $\omega_s$ being (d) $\varepsilon_1$, (e) $\varepsilon_2$, and (f) $\varepsilon_3$, respectively. The red arrow indicates the $m=5^\mathrm{th}$ resonant frequency mode in the rightmost ring is excited.}
\end{figure}

\textbf{Effective magnetic field.}
Photons are neutral particles, but it has been shown that, by introducing the proper distribution of hopping phases in a photonic lattice, one can create the effective magnetic field for photons \cite{51}. In the synthetic honeycomb lattice in Fig.~1(c), we consider the modulation phase as $\phi_{2n-1}=(2n-1)\phi$ and $\phi_{2n}=(2n)\phi$. In each unit cell, the clockwise accumulation of the hopping phase is $-2\phi$, which naturally brings an effective magnetic field. In Fig.~4(a), we consider an infinite synthetic lattice and plot the projected band structure along $\phi$, which gives the butterfly-like spectrum. The choice of phase can be tuned in each modulator arbitrarily. If we set $\phi=\pi/4$, a projected band structure along the $k_f$ axis can be plotted by considering finite number of rings (with $n$$\in$$[1,40]$). As shown in Fig.~4(b), one can see that there are 8 bulk bands, which is consistent with the fact that there are 8 sites in each unit cell once phases with $\phi=\pi/4$ are considered. The middle two bulk bands have degenerate points. Meanwhile, there are 6 gaps between bulk bands, where it supports 8 pairs of topologically-protected edge states. By analyzing the distribution of the eigenstate for each edge state, we can find whether the edge state is located on the left or right edge, as shown in Fig.~4(b). Moreover, the Chern number for each bulk band counting from the highest band can be calculated as 1, 1, $-3$, 2, $-3$, $1$, $1$, respectively.

We then perform simulations in a synthetic honeycomb lattice in 12 pairs of rings with $f$$\in$$[0,11.5\Omega]$. To excite a specific edge state in Fig.~4(b), we choose a single-frequency source field at the frequency $\omega_{m=5,\mathrm{A}}+\omega_s$ near the 5$^\mathrm{th}$ mode with a small detuning $\omega_s$, i.e., $s_3=e^{-i\omega_s t}$ and excite the rightmost ring. The simulations are performed with different parameters and results are plotted in Figs.~4(c)-4(f) at $t=20g^{-1}$. We first set $\phi=0$, so there is no effective magnetic field, and choose $\omega_s=\varepsilon_1=0.9g$. One can see in Fig.~4(c) that the intensity distribution of the field undergoes a random-walk-like propagation in the synthetic honeycomb lattice, and bulk of the lattice is excited. Next we set $\phi=\pi/4$ and introduce the effective magnetic field. In this case, we again consider the excitation $\omega_s=\varepsilon_1$, and plot the result in Fig.~4(d). Different from Fig.~4(c), here one can see the topologically-protected one-way edge state propagating towards lower frequency components at the right boundary, which is consistent with the negative slope of the edge state at $\varepsilon_1$ in Fig.~4(b). We further use $\omega_s=\varepsilon_2=1.7g$ and $\omega_s=\varepsilon_3=2.2g$ to perform simulations and plot results in Figs.~4(e) and 4(f), respectively. In both cases, the excited edge states propagate towards higher frequency components unidirectionally, corresponding to different edge states with the positive slope. Although we study phenomena only associated to the effective magnetic field with $\phi=\pi/4$, the gauge field can be easily tuned in this synthetic honeycomb lattice.

\textbf{Valley-dependent Lorentz force.}
Different from the effective magnetic field for photons introduced by modulation phases, a pseudo magnetic field can also be alternatively generated by applying non-uniform strain in the honeycomb lattice \cite{e4}. In our proposed synthetic honeycomb lattice, we can also easily simulate the effective valley-dependent Lorentz force by varying modulation strength in each ring.

\begin{figure}[htbp]
\centering
\includegraphics[width=0.55\textwidth ]{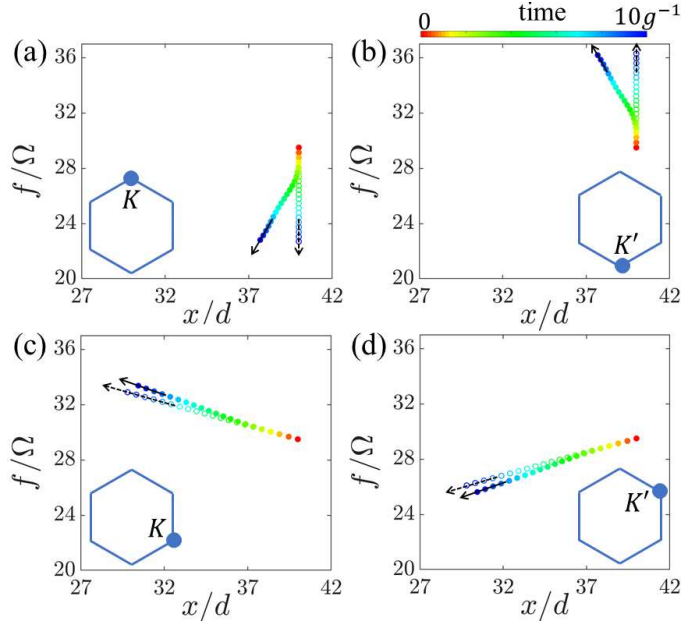}
\caption{\textbf{Valley-dependent Lorentz force.} The simulated trajectories of the center of the Gaussian-shape wave packet $(x_c,f_c)$ at discrete times with time interval $\delta t=0.5 g^{-1}$. The lattice is excited by initial source in the vicinity of the Dirac point $K$ or $K'$ with (a) $(k_x,k_f)=(0,4.4\pi/3\Omega)$, (b) $(k_x,k_f)=(0,-4.4\pi/3\Omega)$, (c) $(k_x,k_f)=(2.2\pi/3\Omega,-2.2\pi/3\Omega)$, and (d) $(k_x,k_f)=(2.2\pi/3\Omega,2.2\pi/3\Omega)$. The trajectory with solid (hollow) circle represents the motion of the wave packet in the case with (without) the valley-dependent Lorentz force, where the moving direction is indicated by the arrow. The varying color of circles represents the center position in the synthetic lattice at different simulation times.}
\end{figure}

We consider a relatively large synthetic lattice composed by 56 pairs of rings with a range of $x$$\in$$[0,83d]$ and $f$$\in$$[0,59.5\Omega]$. We emphasize that, although there are 112 rings considered in simulations for the better illustration, one does not require such a large number of rings to realize the effective valley-dependent Lorentz force in physics. For each modulator, we consider effective modulation strengths having $g_{2n-1}=[1+\eta(3n-42.75)]g$ and $g_{2n}=[1+\eta(3n-41.25)]g$, where $\eta$ is a constant. Following the relation between coupling strengths in a honeycomb lattice and the effective gauge potential \cite{85,86,87}, we obtain an effective gauge potential $A_f(x)\propto g(x)-\kappa$ and $A_x=0$ in the vicinity of the Dirac point, where the relation between $n$ and $x$ is used and $\kappa$ is assumed to be a constant. This effective gauge potential leads to a pseudo magnetic field $B\propto\eta$ and along the $z$ direction. Therefore, one can tune $\eta$ by changing modulations strengths in rings to vary $B$ in the synthetic honeycomb lattice.

In simulations, we inject fields into multiple rings with different frequency components to excite a Gaussian-shape wave packet $s_4=e^{-[(x-x_0)^2/(15.5d)^2+(f-f_3)^2/\Delta^2]}\cdot e^{i[k_x\cdot (x-x_0)+k_f(f-f_3)]}$ in the synthetic lattice, where $\Delta=8.95\Omega$, $x_0=40d$ and $f_3=29.5\Omega$ are the center position of the Gaussian-shape wave packet. The phase information $(k_x,k_f)$ is chosen to excite different states in the vicinity of the Dirac point $K$ or $K'$ in the first Brillouin zone. The simulated motion of the center of the wave packet $(x_c,f_c)$ is then plotted to show the trajectory of the field in the synthetic lattice, with the definition of $(x_c,f_c)$ can be found in Methods.

We first excite the vicinity of Dirac point by a wave packet $s_4$ with $(k_x,k_f)=(0,4.4\pi/3\Omega)$ which gives an initial group velocity pointing towards the negative frequency axis. Without the pseudo magnetic field ($\eta=0$), the wave packet of the field propagates without changing the direction and its trajectory is straight, as shown in Fig.~5(a). Instead, if $\eta=0.004$, the motion of the wave packet is bent to the clockwise side due to the pseudo magnetic field. On the other hand, if we excite the $K'$ point with $(k_x,k_f)=(0,-4.4\pi/3\Omega)$, the motion of the wave packet is bent to the counter clockwise side under pseudo magnetic field in the synthetic lattice [see Fig.~5(b)]. In Figs.~5(c)-5(d), we also excite the vicinity of Dirac points $K$ and $K'$ with $(k_x,k_f)=(2.2\pi/3\Omega,-2.2\pi/3\Omega)$ and $(2.2\pi/3\Omega,2.2\pi/3\Omega)$, respectively. One see that, with a non-zero $\eta$, trajectories of the field are bent towards different directions. The direction of the pseudo magnetic field is dependent on the valley in the synthetic honeycomb lattice, which therefore results in the field bending effect by the effective valley-dependent Lorentz force.

\section*{Discussion}
The highly tunable parameters of modulated ring resonators are of apparent significance in our design for achieving the synthetic honeycomb lattice, which can be realized in potential experiments based on established platforms with fiber loops \cite{51,55,b7,88}, and lithium niobite technologies \cite{b8,6}. For the fiber-based ring resonator, the modulation frequency is $\sim$10 MHz for a fiber length of $\sim$10 m. 2$\times$2 fiber couplers with a high-contract splitting ratio can be used to couple two rings. As for the on-chip lithium niobate device, fields in nearby resonators are coupled through evanescent wave, where the modulation frequency can reach to $\sim$10 GHz when the ring radius is $\sim$2-3 mm \cite{b8}. As an important note, the proposed method that we use to build the synthetic honeycomb lattice through staggered resonances also provides a new perspective for further constructing other complicated lattice structures with C$_3$ symmetry, such as triangular lattice and kagome lattice, both of which hold rich physics in photonics \cite{li2019,schulz2017,Zan2010,struck2011}.

In summary, we use an array of ring resonators composed by two types of rings undergoing dynamic modulations to form a two-dimensional honeycomb lattice in a synthetic space including one spatial dimension and one frequency dimension. We demonstrate a highly reconfigurable synthetic honeycomb lattice which can be used to simulate various phenomena including Klein tunneling, valley-dependent edge states, topological edge states with effective magnetic field, and field bending with the valley-dependent Lorentz force. Our work shows not only the capability for simulating quantum phenomena, valley-dependent physics, and topological states in a modulated ring system, but also points out an alternative way to control the frequency information of light with synthetic dimensions, which potentially enriches quantum simulations of graphene physics with photonic technologies.
\section*{Methods}
\textbf{The construction of the synthetic honeycomb lattice.}
In Fig.~1(a), ring A (B) supports a set of resonant modes with frequency $\omega_{m,\mathrm{A}}=\omega_0+m\Omega$ ($\omega_{m,\mathrm{B}}=\omega_0-\Omega/2+m\Omega$), which is plotted in blue (cyan) color. The effective hopping amplitude between the nearby resonant modes $a$ and $b$ in two rings is formed through a two-step process: the resonant mode $a$ in ring A couples to a corresponding non-resonant mode in ring B via evanescent-wave coupling, and then couples to the resonant mode $b$ in ring B via the dynamic modulation, vice versa. Hence the effective coupling strength $g$ is composed by both the evanescent-wave coupling strength and the modulation strength in EOM \cite{89}, and hence such connections construct a synthetic frequency dimension in a pairs of modulated rings. The construction of effective couplings between resonant modes in a pair of rings with the AB type can also be generalized to the pair of rings with the BA type by mirror symmetry [see the pair of rings labelled by $n=1$ and $\alpha=2$ as an example in Fig.~1(c)]. Therefore, in an array of pairs of rings arranged with alternate combinations AB and BA as shown in Fig.~1(b), the spatially nearby resonant modes at the same frequency can be coupled through the evanescent wave at the coupling strength $\kappa$. Following this procedure, a synthetic honeycomb lattice can be constructed in a space shown in Fig.~1(c) with the longitudinal frequency dimension and the horizontal spatial dimension.

\textbf{Simulation method.}
We expand the field inside each ring as
\begin{eqnarray}
|\psi(t)\rangle =\sum_{n,m}( C_{n,m,\alpha} a^\dagger_{n,m,\alpha} + D_{n,m,\alpha} b^\dagger_{n,m,\alpha}) |0\rangle,
\end{eqnarray}
where $C_{n,m,\alpha}$ and $D_{n,m,\alpha}$ are the field amplitude at the $m^\mathrm{th}$ mode in the corresponding ring. Schr\"{o}dinger equation $i|\dot\psi(t)\rangle = H|\psi(t)\rangle$ is then used in simulations \cite{54}. For different simulations, we take different source $s$ to excite particular rings at specific modes, which we show details in the main text. The excitation process is done by coupling each ring with external waveguides where the light can be injected. The field amplitudes in rings can then also be readout through waveguides. The coupling equation between the external waveguide and the ring is
\begin{eqnarray}
 \begin{pmatrix}
 C_{n,m,\alpha}(D_{n,m,\alpha})(t^+)\\
 E_{\mathrm{C(D)},m,\alpha}(t^+)
 \end{pmatrix}
 =
 \begin{pmatrix}
 \sqrt{1-\gamma^2} & -i\gamma \\
 -i\gamma & \sqrt{1-\gamma^2}
 \end{pmatrix}
 \begin{pmatrix}
C_{n,m,\alpha}(D_{n,m,\alpha})(t^-)\\
 E_{\mathrm{C(D)},m,\alpha}(t^-)
 \end{pmatrix},
\end{eqnarray}
where $\gamma$ is the coupling strength between the external waveguide and the ring through evanescent wave. $E_{\mathrm{C(D)},m,\alpha}(t^-)$ ($E_{\mathrm{C(D)},m,\alpha}(t^+)$) is the corresponding field component injected (collected) through the external waveguide, and $t^\pm=t+0^\pm$.

\textbf{An effective on-site potential induced by frequency shift.}
For a ring, the resonant frequency for the $m^\mathrm{th}$ resonant mode is $\omega_m= \omega_0+m\Omega=\omega_0 +m\cdot 2\pi v_g/L$, where $\omega_0$ is a reference optical frequency and is much larger than $\Omega$ (which is usually in the regime of GHz to THz). Without loss of generality, we can rewrite the equation of resonant frequency as $\omega_m=(M+m)\cdot 2\pi v_g/L$, where $M\gg m$ is a number orders of magnitudes larger than $m$. If we consider the length of the ring changes to $L'=L-\delta L$ ($\delta L\ll L$), the new resonant frequency becomes $\omega'_m = (M+M\cdot \delta L/L +m) \cdot 2\pi v_g/L = (M'+m') \cdot 2\pi v_g/L = \omega'_0 + m'\Omega$ under the first-order approximation, where $m'=m+\mathrm{round}(M\cdot \delta L/L)$ and then $M'$ is shifted by a number smaller than 1. Hence, by changing the length of rings slightly, one is possible to shift the reference frequency $\omega_0$ with a small amount, which results in an effective on-site potential $U \equiv \omega'_0 - \omega_0$ in each ring.

\textbf{Definitions of $x_c$ and $f_c$.}
We define central positions of the Gaussian-shape wave packet in the synthetic lattice as
\begin{eqnarray}
&& x_{\mathrm{c}}=\frac{\sum_{n,m,\alpha} x_{n,m,\alpha}I_{m,n,\alpha}}{\sum_{n,m,\alpha} I_{n,m,\alpha}},\nonumber\\
&& f_{\mathrm{c}}=\frac{\sum_{n,m,\alpha} f_{n,m,\alpha}I_{m,n,\alpha}}{\sum_{n,m,\alpha} I_{n,m,\alpha}},
\end{eqnarray}
where $I_{n,m,\alpha}=|C_{n,m,\alpha}|^2$ or $|D_{n,m,\alpha}|^2$, $x_{n,m,\alpha}$ and $f_{n,m,\alpha}$ are the corresponding position along the $x$-axis and frequency dimension in the synthetic lattice, respectively.
\section*{Acknowledgements}

We greatly thank Prof. Shanhui Fan for fruitful discussions. The research is supported by National Natural Science Foundation of China (11974245), National Key R\&D Program of China (2017YFA0303701), Shanghai Municipal Science and Technology Major Project (2019SHZDZX01), Natural Science Foundation of Shanghai (19ZR1475700), and China Postdoctoral Science Foundation (2020M671090). This work is also partially supported by the Fundamental Research Funds for the Central Universities. L.Y. acknowledges support from the Program for Professor of Special Appointment (Eastern Scholar) at Shanghai Institutions of Higher Learning. X.C. also acknowledges the support from Shandong Quancheng Scholarship (00242019024).

\section*{Contributions}

L.Y. initiated the idea. D.Y. and L.Y. performed simulations. D.Y., G.L., M.X., and L.Y. discussed the results. D.Y., G.L., and L.Y. wrote the draft. All authors revised the manuscript and contributed to scientific discussions of the manuscript. L.Y. and X.C. supervised the project.

\section*{Conflict of interests}
The authors declare that they have no conflict of interest.


\end{document}